# Microfoundations of IPR and standardization strategies of companies: Evidence from the evolving European Single Market

Jussi Heikkilä[1], Satu Rinkinen & Tero Rantala

*Lappeenranta-Lahti University of Technology LUT*

This version: 13 Dec 2024

Abstract

Intellectual property rights (IPR) and standards are important institutions that by shaping appropriability conditions of companies impact international trade flows and the rate and direction of technological progress and innovation activity. We shed light on microfoundations of IPR and standardization capabilities and explore how companies have developed their IPR and standardization strategies and adapted to related institutional changes in the European Single Market. The analysis of the IPR and standardization strategies of companies active in Päijät-Häme region of Finland, a northern part of the European Union, reveals that only a few companies have explicit IPR and standardization strategies, but several have systematic approaches to following the development of standards and IPR environments in their industries. Companies build dynamic IPR and standardization capabilities and adapt their IPR and standardization strategies to the changing institutional environment via experiential learning.

Acknowledgements: Earlier versions of the paper have been presented at the European Regional Science Association (ERSA) Congress 2023 in Alicante/online, LUT University, the 21st Lahti Science Day, Lahti Industry Association (Lahden Teollisuusseura) morning seminar, Evolutionary and historical approaches to organizations seminar at Tampere University and the GEOINNO2024 conference (Uneven Geography of Standards special session) at the University of Manchester. We thank Tuomo Uotila, Pasi Nevalainen, Kerstin Schaefer and conference and seminar participants for helpful comments. Financial support from PHP Säätiö and Business Finland is gratefully acknowledged.

---

[1] Corresponding author: jussi.heikkila@lut.fi.

# 1 Introduction

The importance of intellectual property rights (IPRs)[2] and standards as focal institutions of the international trade system has increased during the past decades (Swann 2010, Blind et al. 2023, Drori et al. 2023). The number of standards impacting the trade of goods and services has increased significantly (Spencer & Temple 2016) concurrently with increasing patent, trademark and design right filings (WIPO 2023). Both evolving IPR and standardization institutions have significant impacts as appropriability conditions (Hurmelinna-Laukkanen & Yang 2022)[3] on how companies can profit from their innovations in global markets (Teece 1986, 2018). The interplay between patenting and standards institutions has been increasingly analyzed (e.g., Blind & Thumm 2004, Grossman et al. 2015, Holgersson et al. 2019, Blind et al. 2022a, 2022b, 2022c, Drori et al. 2023), but there is little empirical evidence on how companies simultaneously develop their IPR and standardization strategies in practice and what are the focal learning events in the process.

Standards may promote trade, but also hinder market entry and cross-border trade or favor local companies (Blind et al. 2018, Blind et al. 2022c) as technical barriers to trade.[4] The impacts depend on the geography of standards – national standards may have different impacts compared to regional and international or global standards and empirical evidence suggests that international standards have the most positive impact on trade (Swann et al. 1996; Swann 2010; Blind et al. 2022c). As global value chains have become complex and economies more interdependent, the role of international standards for international trade partners has become increasingly important (Blind et al. 2018). Schmidt and Steingress (2022) estimated that harmonizing standards[5] have contributed up to 13% of the growth in global trade between 1995 and 2014. However, recent geopolitical developments are creating challenges to the international trade system relying on global standards (cf. Bradford 2020, Zúñiga et al. 2024, Blind 2025).

In this paper, IPR and standardization strategies of companies are viewed via the lens of dynamic capabilities framework (Teece 1997, 2007) with particular focus on their microfoundations (Helfat & Peteraf 2015, Felin et al. 2015) and related (organizational) learning processes (Eisenhardt & Martin 2000, Zollo & Winter 2002). This perspective on dynamic capabilities emphasizes the underlying learning processes in a changing and complex institutional environment resembling the perspective of Zollo and Winter (2002) who define a dynamic capability as "a learned and stable pattern of collective activity through which the organization systematically generates and modifies its operating routines in pursuit of improved effectiveness".

While there is no exact definition for microfoundations in the context of dynamic capabilities (often categorized as sensing, seizing and reconfiguring, cf. Teece 2007, Helfat & Peteraf 2015, Felin et al. 2015), our analysis aims to shed light on the microfoundations underlying companies' current dynamic IPR and standardization capabilities that affect competition and innovation and eventually the rate and direction of technological progress at the macro level. Learning performs a critical role in the development of dynamic capabilities (Eisenhardt & Martin 2000, Zollo & Winter 2002, Vahlne & Johansson 2017). Hence, we are particularly interested in increasing our understanding on the role of companies' stock of accumulated experiences related to IPR and standardization institutions in contributing to evolving appropriability strategies (Teece 1986, 2018; Hurmelinna-Laukkanen & Yang 2022).

---

[2] Throughout the paper we use the term "IPR strategy" instead of "patenting strategy" (cf. Grossman et al. 2015) as IPR is more extensive set of appropriability mechanisms and comprises trademarks, design rights, etc. (e.g., Hall et al. 2014, Heikkilä & Peltoniemi 2023).

[3] Hurmelinna-Laukkanen & Yang (2022, p. 10) define "appropriability conditions as those contextual and situational factors that influence the availability and feasibility of the instruments of appropriability and appropriation processes"

[4] See The World Trade Organization's Technical Barriers to Trade Committee: Principles for the Development of International Standards, Guides and Recommendations, https://www.wto.org/english/tratop_e/tbt_e/principles_standards_tbt_e.htm Accessed 13 Aug 2023

[5] Schmidt and Steingress (2022) use the notion of "harmonized standards" more broadly and do not focus only on European harmonized standards. According to the European Commission: "A harmonised standard is a European standard developed by a recognised European Standards Organisation: CEN, CENELEC, or ETSI. It is created following a request from the European Commission to one of these organisations. Manufacturers, other economic operators, or conformity assessment bodies can use harmonised standards to demonstrate that products, services, or processes comply with relevant EU legislation." Source: https://single-market-economy.ec.europa.eu/single-market/european-standards/harmonised-standards_en Last accessed 27 Nov 2024.

In Europe, European standards have played a focal role in enabling the European Single Market (European Commission 1985, 2022, Pelkmans 1987, 2024, Blind et al., 2018) and Bradford (2020) has coined the term "Brussels effect" that refers to the EU's ability to act as a global regulator and standard setter. The Brussels effect in the context of open European standards can be driven by specific companies who are active in standards development. There are regional differences in how actively local companies participate in developing European standards and how they impact them – who lead (standards makers) and who follow (standards takers) (Tosic et al. 2024). Anecdotal evidence indicates that companies from older EU member countries – that are located closer to Brussels - are more active in standards development compared to companies from more remote peripheral regions. Empirical evidence suggests that participation at standardization meetings is elastic to distance; distance decreases the likelihood to participate onsite (Baron & Rosá 2024).

Hence, there is a need to understand how companies in different parts of the European Single Market (e.g., core vs. periphery based on the distance to Brussels, cf. Spiekermann & Aalbu 2004, Peñalosa & Castaldi 2024) cope with the evolving standardization environment as well as the evolving IPR environment. Established companies and new entrants must adapt to the dynamically changing and increasingly complex institutional environment (Ye et al. 2024) to survive and efficient learning is focal in this continuous adaptation process. How do companies formulate and adapt via experiential learning their IPR and standardization strategies in the European Single Market? How do companies build their dynamic capabilities related to IPRs and standards? How does the changing national, European and global IPR and standardization environment impact businesses of companies?

This study contributes to the emerging empirical IPR and standardization literature (Blind & Thumm 2004, Grossman et al. 2015, Holgersson et al. 2019, Blind et al. 2022a, 2022b, 2022c, Drori et al. 2023) by shedding light on the learning processes of companies in the ever-changing IPR and standardization environments, particularly in the context of the European Single Market. We focus on Finland, providing a view from the perspective of small open economy companies. Firms from small open economies tend to be more likely to internationalize compared to firms from larger home economies (cf. Benito et al. 2003, Heikkilä & Peltoniemi 2023) as the domestic scaling opportunities are limited. The following analysis sheds light on the microfoundations of IPR and standardization capabilities and explore how companies have developed their IPR and standardization strategies and adapted to related institutional changes in the European Single Market via experiential learning.

## 2 IPR and standardization strategies in evolving institutional contexts

According to the Uppsala model of firm internationalization, companies incrementally intensify their internationalization as they accumulate experience via experiential learning (Johanson & Vahlne 1977, Vahlne & Johansson 2017, 2019). Forsgren (2002, p. 273) notes that "as the model is primarily a model about how uncertainty is handled through learning, the concept of learning is of crucial importance." The model has been refined and, more recently, the importance of microfoundations in the context of firm internationalization has been emphasized (Coviello et al. 2017). While IPR systems and standardization systems can be understood as institutions that impact internationalization as well as scaling opportunities and frame the microfoundations of firm internationalization, to our knowledge, the existing literature analyzing their role has received relatively little attention. The institutional environment related to IPRs and standardization in the target markets are crucial (Drori et al. 2023, Tosic et al. 2024).

Globalization and European integration have significantly changed the business and competitive environment of companies (Aghion et al. 2015, Pelkmans 2024). An increasing number of companies are born-globals (Knight & Liesch 2016) – that is, they target and operate in the international markets and global value chains from initiation instead of first focusing on the domestic market and gradually expanding the business to foreign markets. In the context of the European Single Market, the firms are by definition "born EU" companies as their business environment and appropriability conditions are both enabled and constrained by EU regulation as well as European IPR and standards institutions from the start. Next, we provide a brief review of prior literature on IPR and standardization strategies and describe the evolution of the European institutional environment.

## 2.1 IPR and standardization strategies and appropriability conditions

When investing in research, development, and innovation activity (R&D&I), companies need strategies on obtaining return and profit from those investments and innovation activities (Teece 1986, 2018, Hurmelinna-Laukkanen & Yang 2022). We consider both IPR and standardization strategies to be important strategies to appropriate returns from innovation. Hurmelinna-Laukkanen and Yang (2022) define appropriation strategy to be *"a strategic alignment of instruments of appropriability and appropriation processes with each other and with appropriability conditions to benefit from innovation".* Moreover, it should be emphasized that IPR and standardization institutions are not static but evolving – and even uncertain – institutional environments determining appropriability conditions (Hall & Helmers 2019, Prud'homme et al. 2021, Heikkilä & Peltoniemi 2023). Brunsson et al. (2012, p. 627) notes "While standards might aim at the creation of stability and sameness, standardization itself is a highly dynamic phenomenon. Even the stability of standards themselves has to be understood as the result of underlying dynamic processes."

Appropriability literature (Teece 1986, 2018, Hurmelinna-Laukkanen & Yang 2022, Mezzanotti & Simcoe 2023) focuses on analyzing the strategic micro-level choices of companies. While the focus has been particularly on patents, patent statistics and patenting strategies (Hurmelinna-Laukkanen & Yang 2022, Cappelli et al. 2023), IPRs also cover trademarks, copyrights, design rights and trade secrets as well as informal appropriation methods such as lead time (Hall et al. 2014, Mezzanotti & Simcoe 2023). In practice, only a small share of companies relies on patent protection in their innovation activity. For instance, Hall et al. (2014) report that the share of firms patenting among innovating firms is about 4 % in the UK. As a consequence, empirical studies limiting the attention to patent statistics and patent filing activity, necessarily focus on a very limited set of companies. Therefore, it is important to analyze more broadly IPR strategies instead of focusing only on patenting activity. Companies must also have strategies to ensure freedom to operate (FTO) and to manage risks related to infringement of others' IPRs. For instance, pre-emptive patenting – filing patent applications not to obtain protection for the invention but rather to prevent others from patenting the same inventions (Guellec et al. 2012) – is one tool to maintain FTO (Cappelli et al. 2023).

Also, motives (not) to file IPRs may evolve over time as the players in a specific IPR environment learn about the functioning of IPR systems. In other words, the IPR know-how and knowledge about the functioning of IPR institutions accumulates via specific events such as decisions made by the patent office and courts (Heikkilä & Peltoniemi 2019). Heikkilä and Peltoniemi (2019) describe how companies learn the boundaries of design rights sequentially via different events and learning opportunities. In other words, some of uncertainties related to IPRs in a specific industry-context are resolved when their boundaries are tested and clarified during IPR application processes and court cases. This experiential learning process is often both time and resource-consuming and involves significant risks and uncertainties.

The role of external patent attorneys has been increasingly studied, since they may have significant impacts on the companies' capabilities to profit from innovation by impacting the quality of IPRs and IPR strategies (Süzeroğlu-Melchiors et al. 2017, de Rassenfosse et al. 2023, Andriosopoulos et al. (2023), Heikkilä & Peltoniemi 2023). As IPR legislation is reformed, companies may need professional help in analyzing the implications on their businesses. Patent and other IPR attorneys and lawyers play an important role in the learning process and in helping companies navigate the IPR environment (sensing) and exploit the opening opportunities (seizing). This IPR service sector helps companies to adapt to the changing and complex IPR institutions.

Standards exist for different purposes, such as safety, quality, measurement, compatibility and interoperability (Swann 2010), with varying impacts on markets, competition and innovation. Like IPRs, standards codify (technical) knowledge (Blind et al. 2022c), and both can be used strategically as part of business strategies. Therefore, IPR strategies should not be considered separately from standardization strategies. For instance, Grossmann et al. (2015) analyzed the strategic use of patents and standards for new product development knowledge transfer and propose that "standardization activities and the continuous tracking of standards information should be tied to the new product development process in a similar manner as the patenting strategy process".

Blind and Mangelsdorf (2016) studied strategic motives of German manufacturing companies in the electrical engineering and machinery industry to be involved in standards development organizations and found that pursuing specific company interests, solving technical problems, knowledge seeking, influencing regulation, and facilitating market access are important motives. In addition, they found that firms have a particularly strong interest in ensuring industry-friendly design of regulations that can be achieved by standards and that small firms are active in standards alliances to access knowledge from other stakeholders. De Vries et al. (2009) emphasized that trade associations play an important role particularly for small and medium-sized enterprises in keeping them up to date on standardization matters. In the Finnish context, this is also the case as the national standardization system is decentralized meaning that there are specific industry associations that act as standards writing bodies for the national standards body and information sources for member companies (Heikkilä 2024).

Blind et al. (2022a) considered a wider set of potential standardization motives among employees of two German Original Equipment Manufacturers (OEMs) in the automotive sector. The top 5 motives to participate standards development were 1) Help to shape technically mature and industry-oriented standards, 2) Carry forward state of the art technology, 3) Contribute individual abilities for the benefit of the company, 4) Prevention of standards and contents that contradict the interests of the company and 5) Guarantee safety in technology (Blind et al. 2022a, p.8). Many firms that participate in standardization do not themselves contribute much, but instead follow what other companies and entities are developing. Thus, standards development can be an important knowledge diffusion and spillover channel (Blind 2006, Blind & Mangelsdorf 2016).

All companies are impacted by IPR and standards institutions and they must make strategic choices regarding the utilization of IPR systems and participation in standards environment - keeping in mind that ignorance of these institutions is also a strategic choice. Absorptive capacity of companies - that is, their ability to recognise the value of new information, assimilate it and apply it to commercial ends (Cohen & Levinthal, 1990, p. 128) is crucial when companies continuously learn to exploit evolving standards and IPRs.

## 2.2 European integration and evolving institutions of the European Single Market

European integration and the development of the European Single Market are dynamic on-going processes[6], which have had significant impacts on internationalization processes of firms. Figure A.1 in the Appendix illustrates some major changes in the IPR institutions (EU trademarks in 1996, EU registered design rights (RCD) in 2003 and Unitary Patent and Unified Patent Court in 2023) and the increasing variety of harmonized European standards.

---

In particular, from the perspective of companies headquartered in newer EU member states (EU enlargements 1995, 2004, 2007, 2013 and exit of the UK in 2020), the institutional environment was different before and after the EU membership. In other words, learning EU institutions that for non-EU member states were before part of internationalization process, became domestic institutions once the host country became part of the EU. After the EU membership, new firms established in these countries became, by definition, "born EU" firms (cf. "born global", Knight & Liesch 2016). Learning the functioning of European and EU institutions is not anymore learning and adapting to institutions of foreign markets, but instead learning the functioning of domestic markets which complexity has significantly increased. The same applies to internationalization and Europeanization of innovation systems (cf. Carlsson 2006) where international and European IPR and standards institutions have become increasingly important over time (Schmidt & Steingress 2022, Heikkilä & Peltoniemi 2023).

European integration has meant that the importance of European IPR institutions has increased at the cost of national IPR institutions (Hall & Helmers 2019, Heikkilä and Peltoniemi 2023). Similarly, the European integration has progressed in the standardization field and companies from EU member states and elsewhere must design their businesses accordingly and, try to anticipate how the standardization environment evolves in their business environments. The learning processes of companies in the continuously evolving IPR and standardization environments have received limited attention, and this article tries to shed light on this little studied topic.

## 2.3 Research questions

Both evolving IPR and standardization institutions have various impacts on the businesses of companies and on appropriability of their R&D&I investments. They are focal parts of the environmental uncertainty (Haarhaus & Liening 2020) and institutional complexity (Ye et al. 2024) faced by European companies. Companies must develop their IPR and standardization strategies based on their beliefs and expectations regarding the functioning of institutions and strategic choices of competitors as described in Figure 1. Continuous experiential learning ("learning by doing") is inherent in this process of dynamic capability development (Eisenhardt & Martin 2000). Zollo and Winter (2002, p. 348) characterize dynamic capabilities as *"systematic patterns of organizational activity aimed at the generation and adaptation of operating routines"* and propose that these dynamic capabilities *"develop through the coevolution of three mechanisms: tacit accumulation of past experience, knowledge articulation, and knowledge codification processes."* In order to explore the microfoundations of IPR and standardization related dynamic capability development and strategies, we develop the following research questions. It should be emphasized that the institutional context of them is the European Single Market and the perspective is a perspective of small open economy companies.

The first research question describes the status of dynamic capabilities related to IPR and standardization as a stock of accumulated experiences via experiential learning (cf. Johansson & Vahlne 2017). In particular, it can shed light on the extent to which companies are able to articulate and have systematized and codified (Zollo and Winter 2003) their strategic IPR and standardization approaches.

*RQ1. What kind of IPR and standardization strategies do companies have?*

The second research question aims to shed light on the impacts of evolving, uncertain and complex institutional environment (Ye et al. 2024, Heikkilä & Peltoniemi 2023) and appropriability conditions (Teece 1986, 2018; Hurmelinna-Laukkanen & Yang 2022).

*RQ2. What are the impacts of IPR and standardization institutions on companies' businesses?*

The third research question taps into the microfoundations (Helfat & Peteraf 2015, Felin et al. 2015) of IPR and standardization strategies and related dynamic capabilities. Measurement and operationalized empirical analysis of dynamic capabilities (sensing, seizing and transforming, Teece 2007, Teece et al. 2016) is not straightforward (Laaksonen & Peltoniemi 2018, Kump et al. 2019) and, therefore, it is reasonable to focus on the development of IPR and standardization strategies and identifying learning or triggering events. As Teece et al. (2016, p. 18) note: "while strategy and capabilities can be analytically separated, as a practical matter they need to be developed and implemented together".

*RQ3. How do companies develop their IPR and standardization strategies?*

The fourth and final research question is future-oriented and focuses on strategic and corporate foresight (Haarhaus & Liening 2020) by asking the respondents to identify challenges and opportunities related to IPRs and standardization. This is particularly related to the anticipated developments in the evolving institutional environment (incl. European integration).

*RQ4. What kind of challenges and opportunities are related to IPR and standardization?*

**Figure 1. Conceptual framework**

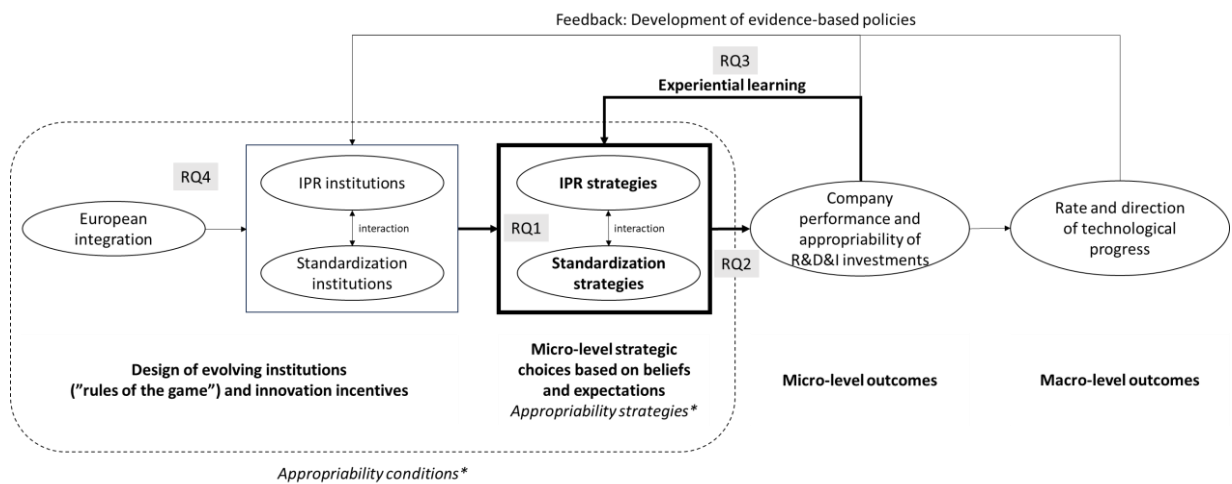

Notes: Authors' illustration, *Hurmelinna-Laukkanen & Yang (2022, p.10) define "appropriability conditions as those contextual and situational factors that influence the availability and feasibility of the instruments of appropriability and appropriation processes" and note that appropriability strategy is "an issue of selecting the optimal [appropriability] mechanisms".

## 3 Method and data

As there is relatively little research on the evolution of IPR and standardization strategies of companies and related learning events, an exploratory case study (Yin 2018) is a justified approach. There is no "paper trail" or pre-existing data to utilize when analyzing the impact of evolving IPR and standardization institutions on those companies' business strategies who do not patent and are not involved in standardization. Therefore, a qualitative approach is required to answer our research questions. The main research method is the analysis of semi-structured interviews which are described in detail in Section 3.2. By inquiring companies directly about their IPR and standardization strategies and related IPR and standardization know-how development efforts, it is possible to learn about their microfoundations and how related dynamic capabilities are developed in practice (Helfat & Peteraf 2015, Felin et al. 2015). What kind of heuristics (Bingham & Eisenhardt 2011, Bingham et al. 2019) and routines (Eisenhardt & Martin 2000, Zollo & Winter 2002) companies have developed via experiential learning, for instance, during their internationalization process (cf. Vahlne & Johanson 2017, Coviello et al. 2017, Niittymies 2020)? Since we focus on companies having activities in a specific region, we provide next a brief description of the local institutional context.

## 3.1 Institutional context

Finland is a small open economy with a population of ca. 5.5 million and GDP of ca. 269 billion euros in 2022. Finland has been a member of the EU since 1995 and a member of the European Patent Convention since 1996. Of 29,000 approved standards in Finland, about 97% were of European or international origin as of 2021[7] which can be interpreted to be evidence of strong European integration. Finland is part of the European Single Market and European standards both enable and restrict business opportunities of Finnish companies. Our study focuses on companies operating in the Päijät-Häme region (NUTS3: FI1C3). Päijät-Häme is an old industrial region in the southern part of Finland and city of Lahti is its capital. The distance from Brussels to Lahti is ca. 1700km, so Päijät-Häme can be considered a Northern peripheral area from the point of view of core regions of the central Europe (cf. Eder 2019, Peñalosa & Castaldi 2024, more details about Päijät-Häme in the Online Appendix).

A company's IPR environment depends on the regional coverage of its business activities and supply chains. In the Finnish context, there are multiple regional "layers" in the IPR environment. There are the national IPR laws and institutions, the European IPR institutions and international treaties where Finland is a member (see Heikkilä & Peltoniemi 2023). For companies operating in the international market, the IPR environment comprises also the IPR environments of target markets other than the European Single Market (cf. Drori et al. 2023). The most important trading partner countries of Finland include Germany, Sweden, the US, the Netherlands and China. Thus, the changes in standards and IPR environments of these countries are focal to many international Finnish companies also including those in Päijät-Häme region.

Since Finland joined the EU in 1995 and subsequently the European Patent Convention in 1996, there has been a shift in IPR filings from national filing channels at the Finnish patent office PRH towards European filings channels, the European Patent Office EPO and the European Union Intellectual Property Office EUIPO (see e.g., Hall & Helmers 2019, Herz & Mejer 2019) and the IPR filing activity of Finnish companies has become increasingly international (Heikkilä & Peltoniemi 2023). EU trademarks (EUTMs) were introduced in 1996 and registered community designs (RCDs) in 2003. They provide EU-wide trademark and design right protection correspondingly and are in force in Finland.

Recently, EU IP Action Plan (European Commission 2020) and standardization strategy (European Commission 2022) were introduced. Also, at national level, the updated IPR strategy of Finland was recently introduced, and the national standardization strategy is on the agenda of the current government as of 2023. These plans and strategies emphasize the awareness and skills related to IPRs and standards. Table 1 illustrates the Europeanization – the shift of focus from national institutions towards common European institutions - of both IPR and standards institutions from the Finnish perspective.

---

[7] https://sfs.fi/en/finnish-standards-association/ Accessed 2 Jan 2023

**Table 1. Trends of Europeanization of IPR and standards institutions from the Finnish perspective**

|  | 1995 | | 2022 | | Change |
| --- | --- | --- | --- | --- | --- |
|  | N | % | N | % | 1995-2022 |
| **Standards** | | | | | |
| Origin of approved standards in Finland* | | | | | |
| National (SFS) | | ~40% | | <5% | - |
| European (EN) | | ~40% | | ~50% | + |
| International (ISO) | | ~20% | | ~40% | + |
| | | | | | |
| **IPRs**** | | | | | |
| Patent grants | | | | | |
| PRH, direct | 2 347 | | 677 | | - |
| PRH, EPO validation in Finland (1996-) | 0 | | 4 473 | | + |
| | | | | | |
| Design right filings | | | | | |
| PRH, direct | 872 | | 68 | | - |
| EUIPO, registered community designs (2003-) | 0 | | 28 783 | | + |
| | | | | | |
| Trademark filings | | | | | |
| PRH, national direct | 6 108 | | 2 655 | | - |
| EUIPO, Community trademarks (1996-) | 0 | | 174 191 | | + |

Notes: *SFS: https://sfs.fi/sfs-ry/meista/sfs-lukuina/ **PRH and EUIPO, PRH: https://www.prh.fi/en/mallioikeudet/tilastoja/mallien_rekisterointihakemukset_ja_haetut_mallit.html (design rights), https://www.prh.fi/en/trademarks/general_information_about_trademarks/trademark_statistics.html (trademarks), https://www.prh.fi/fi/patentit/tilastoja/patentit.html (patents); EUIPO: https://www.euipo.europa.eu/en/about-us/the-office/what-we-do/statistics

## 3.2 Semi-structured interviews

The template of our semi-structured interview is presented in the Appendix. We contacted 32 companies operating in Päijät-Häme region and eventually conducted interviews with 17 companies (18 interviewees as in one case there were two interviewees). Figure 2 illustrates the data collection process and timeline.

We intentionally contacted a set of local companies that are among the oldest and largest in the region, so they have accumulated experience and have had chance to experiential learning for a long period of time about the functioning of both national, European and international IPR and standardization institutions. Majority of the interviewed companies are also export-oriented and have experience of scaling their businesses both within the EU and beyond.

Importantly, these companies have experience about the process of internationalization as described in the Uppsala model (Johanson & Vahlne 1977, Vahlne & Johanson 2017, 2019). All the analyzed companies are already international and have long business histories. Most of these companies were not initially "born EU-companies" but have over time dynamically adapted to the evolving business environment and European integration. This makes them particularly interesting since these companies have had the longest time to accumulate experiences, benefit from experiential learning and develop their dynamic capabilities.

Table A.1 in the Appendix lists characteristics of our 17 interviewees. The average duration of an interview was 42 minutes, and we obtained a total of 11 hours 45 minutes of recorded interview material that were transcribed and analyzed separately by each of the authors (investigator triangulation). Most of the companies are industrial and operate in business-to-business (B2B) type manufacturing industries instead of business-to-consumers (B2C) sectors. Standards may play different roles in manufacturing and service industries, and our sample covers both: 13 (76.4%) manufacturing companies and 4 (23.6%) service companies. The aggregate turnover of the interviewed companies was more than two billion euros in 2022 representing a high share of total revenue of company population in the Päijät-Häme region.

For the sake of transparency, we separate the empirical analysis into two sections. In Section 4, we present the findings and in section 5 we provide our interpretation of the findings and link them to the existing literature.

**Figure 2. Data collection process**

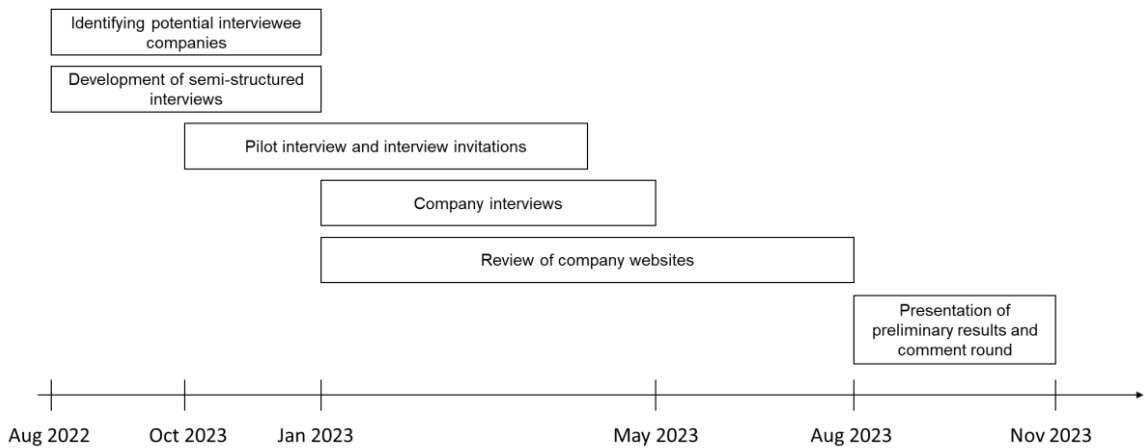

Notes: Authors' illustration

# 4 Findings

## 4.1 IPR and standardization strategies

When asked directly about the existence of a formal IPR or standardization strategy, most interviewees responded that they have neither formal IPR nor standardization strategies with a few exceptions. However, the interviews revealed that most companies had a "systematic approach" (established practices or routines) to IPR and standardization when inquired further. As an extreme example, one company had even been a founding member of specific standards consortium and yet they viewed that they did not have a standardization strategy.

Zollo and Winter (2002) argued that there are three types of knowledge processes that may promote dynamic capabilities. 1. tacit accumulation of past experience, 2. knowledge articulation, and 3. knowledge codification processes. Clearly, the dearth of explicit IPR and standardization processes, but the existence of systematic approaches illustrates how companies have accumulated tacit knowledge about best practices of IPR and standardization which they can articulate to the interviewees, but they less often have codified those approaches. It reflects the process how dynamic capabilities are built upon heuristics (Bingham & Eisenhardt 2011, Bingham et al. 2019). Table 2 illustrates the differences between "codified and tacit" IPR and standardization strategies of our sample firms.

**Table 2. IPR and standardization strategies**

|  | Explicit/Written | Systematic approach* | No systematic approach |
|---|---|---|---|
| **IPR strategy** | 4 | 12 | 1 |
| **Standardization strategy** | 1 | 15 | 1 |

Notes: *Our evaluation on what "systematic approach" is, is based on authors' discretion: if respondents described that they had established practices regarding IPR management and standardization, then we categorized them as having systematic approaches. Note that we did not distinguish here between implementation and development of standards. Representative quotes to which this categorization is based on are available from the authors upon request.

The notions of "IPR strategy" and "standardization strategy" are not unequivocal and can be understood in different ways. If interviewees asked to elaborate what is meant by these concepts, interviewers responded that we intentionally left the definitions open. If companies responded negatively (no IPR or standardization strategy), we asked further whether they had systematic approaches towards the utilization of standards and let them explain their approaches (aim was to minimize confirmation bias). It became clear that it is important to distinguish between A) the use, implementation and monitoring of standards and B) the active development of standards ("standards contributions"). The former "sensing" or monitoring opportunities and threats was part of most companies' business (as usual) whereas only a couple of the companies participated actively in standards development (B) that could be understood as "seizing" opportunities.

Table 3 summarizes the findings and presents representative quotes. In the level of sophistication of IPR and standardization strategies, there seemed to be significant differences. Trademarks were considered very important for business and many companies talked only about the possibility to apply patents. Most companies related IPR first with patenting. However, the patent portfolios of the companies are quite small and play limited role in their businesses.[8] This suggests that IPR portfolios play relatively limited role in seizing the opportunities changing environment offers to the sample companies.

Some companies had owned patents in the past but not anymore. Such IPR strategy change illustrates reconfiguring component of dynamic capabilities. In an industry where product life-cycles are short, unregistered design protection (three years, cf. Andersson et al. 2023) was considered sufficient and, thus, design right registrations were little used.

One company had an extensive documented IPR strategy which main goal was risk management and ensuring FTO in the markets. The main risk management approach was active monitoring of competitors' patenting activity and systematic analysis of patent infringement risks in product development. Patenting was used mainly to promote the market penetration of new products. The patenting decision was based on cost-benefit analysis considering expected economic value of the patent and patent licensing opportunities. Other inventions that were not patented were either intentionally published or kept as trade secrets. Design rights and trademarks were also actively used and counterfeit products continuously monitored. Know-how is protected by non-disclosure agreements as well as R&D contracts. The same company underlined that their own patent portfolio is not very large since patenting costs are fairly high relative to the expected returns although some competitors have quite significant portfolios.

A crucial part of IPR strategies is the strategic make or buy decision: which IP management activities are kept in-house and which are outsourced (Süzeroğlu-Melchiors et al. 2017, Heikkilä & Peltoniemi 2023). Several companies responded that a lot of IP management is outsourced to specialized IPR attorney firms. They also emphasized that the outsourcing to specific patent attorney firms was based on long-term relations that sometimes have lasted for decades. The companies utilizing patent attorney firms were further asked how they decide on the division of labor. IPR landscape monitoring and information search as well as drafting patents was often outsourced to professional patent attorney companies. Recently, some companies have increasingly adopted IPR software for monitoring the IPR environment and competitors' IPR activities (sensing).

The systematic approaches and practices that the respondents have regarding standardization can be summarized as follows: 1) implementation of existing standards, 2) monitoring of standards development and anticipation of future standards and 3) active participation in standards development with explicit agenda. Descriptions of "unwritten" or undocumented standardization strategies ("systematic approaches") revealed that monitoring of standards development and particularly what larger players do, has an important role. The most sophisticated companies had dedicated persons by business units who follow the evolution of relevant standards. The important role of industry associations as standardization information intermediaries is discussed in Section 4.3.

---

[8] Standard-essential patents are particularly important in the information and communications technology industry (cf. Holgersson et al. 2018, Bekkers et al. 2023), but the interviewees did not have experiences of standard essential patent licensing.

Our observations are consistent with Blind and Mangelsdorf (2016) who found that in the context of German electric engineering and machinery industries an important motive to participate in standardization was "the interest in ensuring industry-friendly design of regulations". Most companies stated that they participate in standards development to ensure that no such requirements are included in the standards that could undermine their business.

Finally, the geographical dimension or the regional coverage of IPR and standardization strategies was touched upon in several interviews. Roughly, this dimension could be reduced to A) international, B) European and C) Finnish (national) factors. Generally, in the context of standardization strategies the focus was at EU-level. In addition, international management standards ISO9001, ISO14001, ISO45001 and ISO27001 were mentioned by several companies (more in Section 4.5).[9] In contrast, in the context of IPRs, the respondents focused on IPR protection in specific target market countries. This reflects the fact that IPRs are enforced at national level. The European Single Market is clearly the most important target market of all sample companies.

As companies have internationalized beyond Finnish borders, they have typically had a specific pattern or sequence in their internationalization activities reflecting the gradual internationalization process described by the Uppsala model (Johanson & Vahlne 1977, Vahlne & Johanson 2017, 2019). The responding firms had had different internationalization paths, but some exporting markets were more commonly mentioned including Nordic countries and Germany. Since Finland joined the EU in 1995, companies have been operating in the European Single Market which has made internationalization and scaling much easier. European integration related to standards and the increasing number of European harmonized standards seem to impact heavily the businesses of local companies while IPR systems less so.

## 4.2 Impacts on business

The interviewees' responses revealed that IPR, both their own and competitors', had quite modest impacts on their businesses. Thus, the role of IPRs in seizing opportunities is relatively limited. Patents and trademarks were among the most important protection methods whereas design rights and utility models were mentioned only by a few interviewees.

None of the companies said that their patents totally prevent competitors from entering specific markets but rather that patents slow down their market entry. One company had the experience of regional differences that slowing down of imitation by competitors is easier in the western markets compared to the Asian markets. While patents were not viewed to provide significant competitive advantage in most cases, their role as bargaining chips in negotiations with competitors was acknowledged. No expert highlighted the importance of utility models ("petty patents") that are available in Finland. However, one company noted that they apply utility models in China since they provide quick protection and are challenging to invalidate.

Trademarks were mentioned as important means to protect the brand of the company by almost all interviewees. Typically, brands protected by trademarks are viewed to be more important among B2C (consumer businesses) companies, but the observations indicate that they are the most important IPR among B2B companies as well. Those few companies that mentioned design rights had relatively negative views regarding their capability to exclude competitors from copying the protected designs in line with prior research by Heikkilä and Peltoniemi (2019). On the other hand, there were some exceptions. Chinese design right system was viewed to function well by one interviewee.

---

[9] On its website, SFS reports that ISO 9001, ISO 14001, ISO 45001 and ISO/IEC 27001 were the most purchased standards documents in addition to SFS 6002 and SFS 6000 (electrical safety). See https://sfs.fi/sfs-ry/meista/sfs-lukuina/ Last accessed 24 Nov 2024.

We extended the discussion from registered IPRs also to non-registered IPRs (cf. Hall et al. 2014), including trade secrets and data. Both were considered quite important regardless of industry, particularly continuously accumulating data. Data ownership and interfaces were also discussed in some of the interviews and cruciality of data ownership negotiations were emphasized by some companies. Data has increased its importance in some companies' businesses and predictive maintenance was mentioned as one example where data collection from customers is crucial. One company also mentioned domains as an important category of IPRs but added that preventing cybersquatting[10] is an "endless swamp".

The quotes in Table 3 illustrate how EU regulation heavily impacts the business environment of sample companies – both enabling and constraining their activities and strategic business and R&D&I choices. If a company satisfies the most stringent EU requirements, then it is easy to access other markets such as the US and China where requirements can be less stringent. National standards have decreased in importance as the shift in standards has been to the EU-level. There are also some negative aspects that companies identified regarding regional standards: Fragmented national standards hinder market entry and access (of new innovative solutions) which frustrated some of the companies.

The impacts of IPR and standards on the companies' businesses varied greatly by industry as expected. Some companies noted that their whole businesses are built upon standards whereas others (e.g., textile industry) said that standards have very little impact. In certain industries, standards (and customer requirements related to standards) were the basis of the business and provided the boundaries for R&D activities. An expected finding was that standardization was considered to enable scaling of business. European standards are so stringent that they typically satisfy requirements beyond Europe as well, for instance in Middle East. This can be viewed as evidence of the Brussels' effect (Bradford 2015, 2020) in action.

Complementary, fragmentation was the clearest concern as well as complexity related to keeping up with the increasing amount of EU standardization and regulation. Compliance costs increase as the amount of regulation increases (cf. Maskus et al. 2005). For instance, some companies noted that there are national standards in the built environment which destroys the scale effects. The respondents had contradictory views regarding "upward harmonization" (see Bradford 2020, Vogel 1997 and "California effect") – that is, harmonizing standards within the Single Market to increasingly demanding ones and not the minimum requirements. Some saw it as positive phenomenon whereas others viewed that too stringent EU requirements can hinder competitiveness of local companies. In many companies' business, customer requirements regarding standards were the driver of their own standardization efforts. This derived demand for standards is analogous to the general derived demand in the B2B context: end-customer demand impacts the derived demand in the supply chain. The end-user customer requirements must be taken into account upstream the supply chain.

Certificates for ISO management standards were seen important.[11] Cybersecurity standards - ISO/IEC 27001 specifically - were considered to impact significantly businesses of some companies. IEC 62443 Industrial cybersecurity was also mentioned by one company. Finally, an aspect raised by some interviewees was the impact of chemical regulation, in particular Regulation concerning the Registration, Evaluation, Authorisation and Restriction of Chemicals REACH (and SCIP[12], cf. Bradford 2020).[13] One company considered REACH to be the most important "standard" that impacts its business and a couple of other interviewees mentioned chemical regulation as well.

---

[10] Cybersquatting refers to the practice of registering Internet domain names with the intent to benefit from the goodwill of others' brands and/or trademarks.

[11] Consistent with SFS:n own announcement that management standards are their "best-sellers". See https://sfs.fi/sfs-ry/meista/sfs-lukuina Last accessed 22 Nov 2024.

## 4.3 Learning paths and development of dynamic IPR and standardization capabilities

The most common answer regarding the learning paths in the context of IPR management was that learning happens via doing. As noted in Section 4.1, several companies relied on the advice of professional patent attorney firms and occasionally participated in trainings provided by them (e.g., the impacts of the Unitary Patent system). One interviewee told that their company had invested in IPR contracting skills by hiring a specific in-house legal counsel to be in charge of these issues and educating other employees about IPRs. Some companies described how their R&D workers have benefitted from analyzing patent landscapes – that is, "learning by reading patents". Patent software tools got some mentions. Some companies described how they had witnessed their competitors obtaining patents for inventions that they thought to be obvious and non-novel. These experiences had changed their perspectives towards patents. Another company also pointed out that it is important to own patents since, according to their experience, it seemed that those companies that do not own patents are more likely to become targets of patent assertion in the US.

Learning by doing and by following own industry were considered the primary approaches to acquiring information about relevant standards and developing standardization approaches by multiple companies. Companies accumulate experience and employees learn during customer projects. Some companies participated in standards-related trainings, and it was noted that partners, auditors, and industry associations play important roles in promoting standards awareness and know-how. In addition, some customers invite their suppliers' employees to participate their own education programs regarding standards.

Three of the interviewed companies explicitly mentioned SFS Online service – the online collection of standards offered by the Finnish Standards Association SFS. Learning by reading standards was also one of the mentioned sources of capability development. This is consistent with Spencer and Temple (2016) who argue that standardization constitutes an important mechanism for learning. In addition, industry associations were mentioned as important information sources regarding standardization (cf. de Vries 2009). Due to the small sample, we do not provide a list of mentioned industry associations as that would compromise the anonymity of our interviewees.

Several interviewees noted that in their companies one or a couple of experts are following the standards environment aside their other duties. One larger company told to have "a norm engineer" who solely focuses on standards. Hence, organizational capabilities to sense and seize standardization opportunities are the responsibility of these persons. Generally, there seemed to be a trend from reactive to more proactive standardization approach. One company described how their approach regarding standards has evolved from reactive towards more systematic and proactive approach via learning. At some stage, they had been late in implementing certain standard which had led to challenges with market access. Thus, certain learning events can catalyze more systematic and strategic approaches towards standardization. Companies not systematically following standards were clearly in minority and participation in standards development was considered important. However, as a small player, you must "choose your battles" wisely as noted by one interviewee. In-house R&D capabilities are scarce, and employees participating in standards development are often the best R&D employees, so there is a trade-off between allocating their effort to internal R&D or collaborative standards development.

---

[12] SCIP is the database for information on Substances of Concern In articles as such or in complex objects (Products) established under the Waste Framework Directive (2008/98/EC). Source: https://echa.europa.eu/consumers-and-scip Accessed 1 Aug 2023.
[13] https://echa.europa.eu/consumers-and-scip Accessed 1 Aug 2023.

While most of the interviewees had joined their companies after Finland joined the EU in 1995, several of them still regarded the EU membership and European integration as a crucial process that subsequently have significantly impacted their businesses mainly positively. Interestingly, some companies had contradicting views on the impact of Finland's EU membership on the standardization environment. When asked about the countries where the main competitors active in standards development operate, particularly Middle-European countries and companies headquartered in Germany, France and the Netherlands got mentions. This supports the view of the companies from the old EU countries and core regions as focal standards makers in contrast to the companies from newer and more peripheral EU countries as standards takers.

## 4.4 Future challenges and opportunities

The evolving institutional environment frames the appropriability conditions (cf. Hurmelinna-Laukkanen & 2022) and its changes offer both challenges and opportunities for firms depending on their current position. At the time of the interviews (spring 2023), it was expected that the Unitary Patent system would begin in June 2023 which eventually happened. In some interviews the topic was mentioned, but none of the firms saw it as crucial from their business perspective. Cost savings and simpler filing process enabled by the Unitary Patent were viewed as a positive impact of the new institution. However, lack of resources was clearly an obstacle to more extensive utilization of IPRs.

While many companies owned considerable IPR portfolios, most still viewed the enforcement of IPRs very costly (esp. in the US) and comprehensive monitoring of IPR infringements nearly impossible (esp. in Asia). On the other hand, patents were viewed as part of marketing strategy by some. An important challenge raised by several interviews was the maintenance of FTO: how to ensure that business is not infringing others' and competitors' IPR? Some respondents had been surprised to find out how some obvious non-novel inventions had been granted patents. Thus, some companies file patent applications to prevent other companies from obtaining patents for similar inventions. This approach is sometimes called "pre-emptive patenting" (Guellec et al. 2012, Cappelli et al. 2023). Another company described that a key challenge regarding IPR strategy is to evaluate case by case whether it is better to protect, keep secret or publish (to prevent others from protecting).

Some interviewees openly expressed frustration regarding the challenges posed by standards-related trade politics. Generally, the main benefit of standards was seen as enabling scaling while the main downside was considered too strict requirements and proliferation of standards which increases compliance costs and makes market access and entry more challenging for new players. The companies involved actively in standardization noted that their main participation motivation is to try to keep impossible (product) requirements out of the standards text.

Sustainability standards were mentioned in several interviews and an identified challenge is that there are many different sustainability certificates (e.g., Ecovadis medals). Some interviewees mentioned that client companies' (typically global multinational companies) codes of conducts refer to standards and they as a supplier must comply with the code of conduct. These were viewed to increase the burden and costs of suppliers. Generally, companies viewed the tightening of standards as the trend in sustainability reporting. This requires increasingly resources and increases compliance costs.

Some companies had had negative experiences regarding standards and viewed the whole standardization ecosystem as a playing field for dominating larger players where incumbents – who are often the ones impacting their contents – can use standards as tools to make it harder for competitors to enter the market. On the other hand, companies had also positive experiences and noted that active participation to standardization may well pay-off. But monitoring is not enough, you must actively participate in standardization work then. Slowness was several times mentioned as a challenge regarding standardization and certification processes.

**Table 3. Summary of findings**

**a.**

| | IPR | Standardization |
|---|---|---|
| What kind of IPR and standardization strategies do companies have? | **Most companies do not have explicit strategies but rather varying systematic approaches** | |
| | • *"We've developed strategic work, but there is neither written IPR nor standardization strategy. But it can be said that we do have a clear unwritten IPR strategy at the moment and not necessarily a standardization strategy, but procedures for systematic standardization, or to comply with the standards."*<br>• *"One could perhaps say that we have IPR strategies in each of our three business areas. They are different because the business areas are slightly different in nature. We have not developed an IPR strategy at the group level, though we could do that too, and in that we could still acknowledge and identify the differences that exist in these different business areas. The standardization strategy is certainly not written down. But there is a certain practice and idea about the role of standardization. It can be discussed further, but at least I don't know that it would have been documented in the same way."* | |
| | **Large variation in strategic IPR approaches** | **Companies follow what larger competitors are doing** |
| | • *"In short, our strategy is that when it makes sense, we can patent."*<br>• *"Our policy is simply that we try to find out if there are any [IPRs] that could possibly limit our operating environment, and on the other hand, we aim to ensure that if it is possible to protect, then we try to protect ."*<br>• *"The IPR strategy means that all physical appearance is protected by design patents or trademarks. And then such differentiating factors of the product that are patentable and are not externally visible. - that is, if secrecy is out of question- then they are protected by patents."* | • *" - - an unwritten strategy seems to be that we monitor pretty closely what the big players are doing, so that we can be involved in the right way. That you don't put money into something that won't become a standard."*<br>• *" - - compared to these Central European industry players. In that sense, we don't have much resources to invest in active participation [in standardization]. So, it's more that we try to actively follow what happens and avoid surprises."* |
| | **IPR management and landscape monitoring is often outsourced to professional patent attorneys with which trust-based relationships have been built over long periods of time.** | **Some companies have designated employees whose focal task is to monitor the development of standards and participate standards development** |
| | • *"We use an external patent attorney. In each product development project we conduct IPR search. She conducts the searches in different databases. We go through the findings with the experts. Also, when new ideas are generated in projects, we check their patentability. We frequently monitor what kind of new patents have been applied and go through them with the patent attorney a few times a year."*<br>• *"We have oursourced the management of our trademark portfolio to [X], who have managed our IPR portfolio for decades. They monitor if there are [trademark] registration that could be confused with ours - - ."* | • *"We have a designated person for each business unit who follows those standards, because we have an obligation to follow their new versions. Kind of a contractual obligation to follow their versions and ensure that those products meet those standards."*<br>• *"We have seen in recent years, a few years ago, that with active participation from here in Finland, you can have an impact in the same way as these big countries do. For many years we had our own person in one of the technical working groups, and he also wrote key parts for both standards."* |
| What are the impacts of IPR and standardization institutions on companies' businesses? | **Trademarks and patents are important, design rights and utility models less so. Protection of data has increased in importance.** | **Standards are crucial for several companies** |
| | • *"Of course, trademarks are self-evident. Protecting the brand is very important."*<br>• *"If you want proper protection and enforce your rights, then you must have a patent."*<br>• *"Designs - - they are quite weak."*<br>• *"We have not considered utility models to be particularly attractive. Its protection is afterall quite weak according to our experience."*<br>• *"What we protect is who can use that data and with whose permission."* | • *"Our product development unit must know these standards by heart."*<br>• *"We live from standards. - - We are takers in this regard. - - For us, standards are a lifeline - - It doesn't make sense to develop standards ourselves, but we have to make use of existing standards."*<br>• *"It is the very basis of our business, that all products are made according to the standard, this way they meet the standard requirements, which is needed [in this field]."*<br>• *"Like I said, a bit like a license to operate, we must have those certain certificates and there is quite a long list of them. And since we are a significant player in our field also globally, we also participate in the development of certain certificates that directly apply to these products of ours."* |
| | **IPRs have modest impacts on businesses or the impact is challenging to quantify** | **EU and the Brussels effect have big impacts** |
| | • *"In my opinion, they [IPRs] don't have direct impacts. It's indirect. - - the impact on revenue is really challenging to estimate."* | • *"National standards don't really mean a thing in this field anymore - - . In practice, we play at the European level."*<br>• *"It has been learned in this field from the school, what the role of standards is, and how they should be used. But really, that big change has actually happened here, in this field too, with the EU."*<br>• *"The good thing is that when you make a product that meets that standard, you can sell it all over Europe and most of the Middle East."* |

b.

| | IPR | Standardization |
|---|---|---|
| *How do companies develop their IPR and standardization strategies?* | **Learning by doing (via customer projects) is the basis.**<br>• *"It's a continuous process. We learn more in each project. And in practice, even new designers have to familiarize themselves with patents or patenting."* | **Learning by reading standards is important**<br>• *"Now, for example, ISO 27000 [cybersecurity] has been used in training. Quite a lot of free web courses and other such things where the crowd goes. And then just reading the standards, that's how you learn."*<br>• *"- - we have SFS online, they monitor all the time, if there are any changes, we get a new message about it, in a way always have updated standards available."* |
| | **Learning by reading patents, patent landscaping.**<br>• *"We have invested in our own legal person, who specializes in contracts. He trains product organizations and also sales. Awareness of the subject is managed through this. We have organized, for example, procurement training. If we consider others' contract terms, what we must be able to read and interpret about IPR."* | **Trade associations are important information sources**<br>• *"Yes, we have these key standards according to which products are made, and we have such monitoring of them, so that if something new appears that is not directly monitored, then we also get information. And of course, the different associations we are involved in, like in China for example the [X] association and in the US [X] association and in Europe there is now the European [X] organization, so that's also where we get this current information and more material about the development of standards that is utilized."*<br>• *"We receive requests for comments through [X industry association], that is, through a national organization - - I know these people who are active there. We can comment if there is a comment round, and we've done that."* |
| | **Purchasing services from patent attorney firms**<br>• *"We use an external patent attorney."*<br>• *"We have own patent engineer but we also use an external patent attorney firm."*<br>• *"Yes, we had a training organized by our patent partner a couple of years ago focusing on how one identifies these [IPRs], applies them and what are the requirements for patents, design rights and others."*<br>• *"For some time, we used purchased patent [landscaping] tools, which enabled us to map the market and what is being patented in our area. But now the same thing is bought as a service from these external actors, that is, how [the landscape] is developing. And then we go to current trainings every now and then to get updates. But that patent monitoring is practically outsourced at the moment ."* | **Shift from reactive to proactive standardization strategy**<br>• *"In Europe, the development is relatively slow. - - But it affects us, if we have to make some change to the entire product family and then certify that change, then it's quite a big project, every time, and may cause product discontinuations and others as well. Getting foreknowledge is really important, so you know what should be done and when. - - our systemic approach has developed and improved because at some point we had some tight situations with late reaction and we were even late, so that there became market obstacles because we had not reacted in time."*<br>• *"- - yes, we clearly have the will to be more systematic here, in developing our own strategy. - - Where we want to be involved and where we need to focus on? But how it materializes in practice... in that sense, we are still a pretty small organization, as far as this team is concerned, when you compare, for example, with these Central European industry players. There aren't a lot of resources to put into it [standardization] to participate so actively. Maybe it's more that you try to actively follow what's going on so that there are no surprises afterwards, like that you should have paid attention to this and then it has to be done in a rush."* |
| *What kind of challenges and opportunities are related to IPR and standardization?* | **Obtaining protection and enforcing IPR is costly**<br>• *"As I said, this most often comes down to a lack of resources, but then if you don't take advantage of it [IPR protection], then you also lose money, when you make an innovation and you didn't protect it, so that's a loss then."*<br>• *"But then it's also very expensive and laborious to follow it, when you've protected something. To follow how is it used somewhere? Especially if the market and the country are the kind, that you're not necessarily even there. And indeed, starting processes outside of Europe, that's [expensive and laborious]. You have to really consider it, because there you are, in such an uncertain operating environment, forced into a potentially very laborious and expensive process."*<br>• *"There is always a big financial risk in challenging that the patent should not have been granted. If it's in the US, for example, then there are 100,000 or millions of people who want to go to court, and it's a bit of a lottery to see who will win."* | **Proliferation of standards and more strict ("must have") standards increase compliance costs and entry barriers**<br>• *"Well, maybe the challenge from a business perspective like this is that there are so many [standards]. Just managing the various audits and auditors, we have some certificates that no one in Finland is qualified to audit. And then the question always arises that when this demands quite a lot from the entire organization, then when these become a bit like "must-have" will we still get added value from them? It's such a time-consuming and resource challenge how to handle this whole."*<br>• *"Their number is increasing all the time. They have 'an employment effect'. Doing business in that way increases costs, and they are a bit like "must have". There needs to be a balance all the time, that where we must invest in at this point."*<br>• *"There are a lot of these things, which everyone seems to have, but then the more we go to the slightly more complicated and demanding level, then they are also a bit of an obstacle to entering the market, if a competitor suddenly decides that I'm going to start doing [X], so it's not terribly easy to get there. It's precisely the fact that when things are pretty tightly staked, you can't accelerate from zero to hundred instantly."*<br>• *"Of course, to an increasing extent, these management standards are also required when we are talking about something like larger tenders. Companies and public administration organizations may require them in their own tenders."* |
| | **Freedom to operate must be ensured**<br>• *"As I already mentioned, the number of patents. Some of our competitors, they have a shocking amount. The fact that we really have the certainty that we are not infringing on a patent requires a lot of time and a lot of experts need to be used to go through it. It is clearly a challenge in terms of the project's time management, to get the certainty that there is this freedom to operate."* | **Monitoring is not enough, contribution matters.**<br>• *"We have seen in recent years, a few years ago, that with active participation from here in Finland, you can have an impact in the same way as these big countries do. For many years we had our own person in one of the technical working groups, and he also wrote key parts for both standards. So yes, you need activity. Monitoring alone is not enough."*<br>• *"Above all, the goal is for the standards to be such that they do not cause insurmountable challenges for us. That it is possible to operate. So that the standards are not unreasonable."* |
| | **Varying expectations of the new Unitary Patent System and Unified Patent Court**<br>• *"It is going to exactly the right direction for us. In principle, this will simplify these practical procedures little by little - - European procedures are good, and they will be supplemented and improved. It's a very positive development for us ."*<br>• *"It does not impact us at all."* | **Sustainability and cybersecurity standards are increasingly important**<br>• *"Yes, it's clearly visible that the requirements are getting tighter all the time, and the industry is a green industry, and that's how companies want to act responsibly and all kinds of surveys are filled out. They are not based on standards, but on various processes of large companies."*<br>• *"Well, the challenge is that the number of requirements is growing at an incredible rate at the moment, and it comes specifically from cybersecurity, sustainability, all material certificates, all that kind of stuff, it creates an incredible amount of additional work, which is then difficult to transfer to the prices this additional fixed cost. And on the other hand, it is a prerequisite to be able to do business in the field. And then again, it's related to the opportunity that you do it better than a competitor, for example with sustainability, and then you win deals."* |

Notes: Selected representative quotes. Translated from Finnish interviews.

# 5 Interpretation of findings and discussion

## 5.1 Microfoundations of dynamic IPR and standardization capabilities

Our rich explorative qualitative interview data sheds light on companies' strategic choices and microfoundations of dynamic capabilities (Teece 2007, Helfat & Peteraf 2015, Felin et al. 2015) related to IPR and standardization strategies in a specific institutional and regional context characterized by specific appropriability conditions (Teece 1986, 2018; Hurmelinna-Laukkanen & Yang, 2022). In other words, how companies have adapted to an evolving and complex institutional environment (Ye et al. 2024) and continue to build IPR and standardization related dynamic capabilities via experiential learning (Zollo & Winter 2002, Vahlne & Johanson 2017, Coviello et al. 2017, Niittymies 2020). The particular focus is on the European integration since the shift from national IPR and standards institutions towards European and international ones has been a trend during past decades (Hall & Helmers 2019, Heikkilä & Peltoniemi 2023).

A growing strand of literature analyses the interplay between patenting and standards institutions (e.g., Blind & Thumm 2004, Grossman et al. 2015, Holgersson et al. 2019, Blind et al. 2022a, 2022b, 2022c, Drori et al. 2023). Our observations show that both patents and standards impact the businesses and R&D activities of companies: most interviewed companies reported that they follow both IPR and standards landscape and tune their R&D trajectories accordingly. The interviews provided a variety of perspectives on how companies adapt their businesses and strategies to the evolving IPR and standardization environments. The learning paths are of particular interest as other companies can learn from past successes and failures. We find that only a few of the companies have explicit IPR and standardization strategies, but several have systematic approaches to following the development of IPRs and standards in their industry (Table 2). These mentioned systematic approaches could be considered as heuristics or rules of thumb (cf. Bingham & Eisenhardt 2011, Bingham et al. 2019).

Patent attorney firms were found to play important roles as part of companies' IPR strategies consistent with recent studies (Süzeroğlu-Melchiors et al. 2017, de Rassenfosse et al. 2023, Heikkilä & Peltoniemi 2023). Almost all the interviewed companies relied on professional IPR attorneys and outsourced significant parts of their IPR management activities. Hence, companies' dynamic capabilities, the capabilities to sense, seize and reconfigure IPR-related opportunities rely heavily on outsourced expertise. In these long-term trusted relationships with partner patent attorney firms, the companies often have a named IPR professional, such as patent attorney, that interacts with the company and its managers. However, when asked about the financial impact of IPRs, many companies responded that registered IPRs, mainly patents and trademarks, play only a minor role in their revenue generation.

Several interviewees noted that the European integration, European standards and extension of the EU have promoted their exports and international scaling of businesses. Hence, they have mainly benefitted from the European integration. On the other hand, in the industrial B2B market, the demand and requirements of customers are important drivers in the adoption of standards and some interviewees viewed concerns that the demand for different certificates for compliance with standards has increased significantly during the past decade. Getting and maintaining the certificates with frequent auditing costs directs resources away from other important business functions. Figure 3 illustrates the microfoundations view on the evolution of IPR and standardization strategies and Table 4 presents examples of specific actions and learning events that are related to the microfoundations of dynamic IPR and standardization capabilities.

**Figure 3. Dynamic IPR and standardization capabilities and strategies in a changing institutional environment**

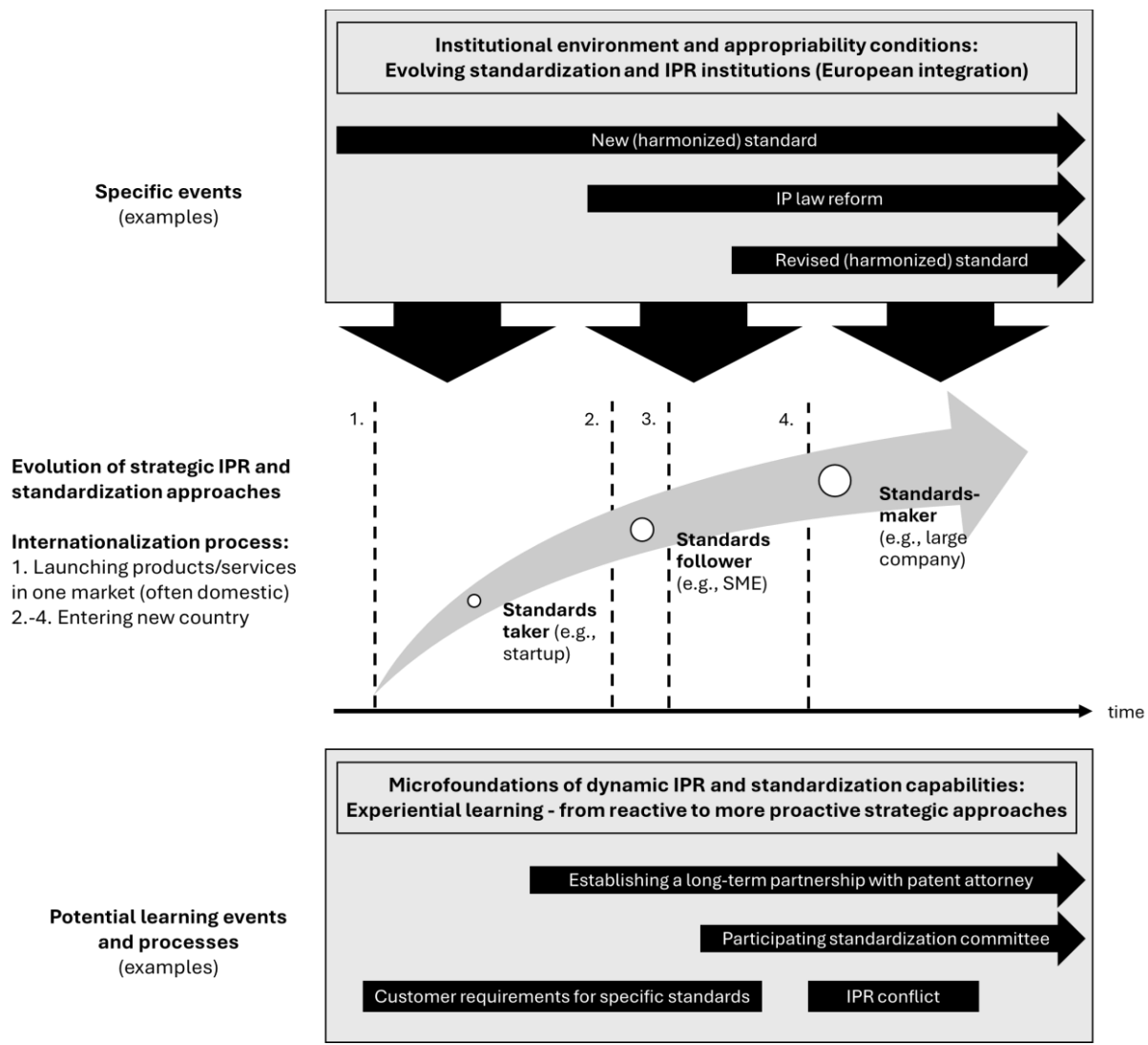

Notes: Authors' illustration

**Table 4. Microfoundations of dynamic IPR and standardization capabilities**

| Microfoundations of dynamic capabilities | IPR | Standardization |
|---|---|---|
| **Sensing** | Action: Monitoring IPR environment, patent landscaping, competitive intelligence | Action: Monitoring standardization environment, foresight |
| | Learning event/process: Advice from patent attorney firm | Learning event/process: Customer requirements, information from industry association, SFS online |
| **Seizing** | Action: Filings and managing IPRs, enforcement of IPRs, IPR disputes | Action: Participating standardization committee, certificate for management standard (ISO 9001, ISO 14001) |
| | Learning event/process: Advice from patent attorney firm, IPR disputes | Learning event/process: Standardization committee meetings |
| **Reconfiguring** | Action: Hiring (firing) a patent engineer | Action: Contributing to standardization (standards maker), investing resources to "norm engineer" |
| | Learning event/process: IPR disputes | Learning event/process: Preparing for new harmonized standards |

Notes: Authors' illustration

Lack of resources is known to be an important obstacle to participate in the development of standards (e.g., de Vries et al. 2009, Blind et al. 2022b). There is a trade-off, whether R&D people participate standardization efforts (public good) or whether they develop products. While the interviewed companies were among the largest in Päijät-Häme region, many considered themselves as relatively small players in the European and global standardization contexts. In other words, they did not view themselves as important standard-setters or drivers of the Brussels effect with a few exceptions. This is additional empirical evidence indicating that smaller players can be in a weaker position to impact the development of standards in line with earlier studies (Blind & Mangelsdorf 2016).

## 5.2 European integration, scaling and dynamic capabilities

There exists a broad literature studying firm internationalization and Uppsala model is one theoretical framework to describe this gradual process where companies incrementally expand to international markets and accumulate knowledge about them (Johanson & Vahlne 1977, Vahlne & Johanson 2017, 2019; Coviello et al. 2017). When company internationalizes, it must learn the IPR and standardization environments and other appropriability conditions (Hurmelinna-Laukkanen & Yang 2022) in the target markets and the more it invests in learning and, for instance, designing the products to conform to the national standards, the more it shows "market commitment". Similarly, protecting inventions with patents, brands with trademarks and designs with industrial designs in specific target markets signals market commitment to those specific markets.

However, learning the "rules of the game" in the target market is not enough as the laws, regulations, standards and the IPR landscape (e.g., who owns which patents and other IPRs) change continuously. There is a constant need to monitor the evolution of the institutional environments, competitors' IPR portfolios and standardization activities as well as manage own IPR portfolio and standardization efforts. The learning events and efforts related to the development of these capabilities are, in essence, microfoundations of dynamic capabilities and successful IPR and standardization strategies.

European integration and the development of the European Single Market are dynamic and evolutionary processes where companies need to adapt to the changing complex and uncertain environment. This requires learning capabilities, and the learning occurs via specific learning events and processes (Figure 3). While European integration has progressed, the process – both from the perspective of standards and IPRs (Hall & Helmers 2019, Heikkilä & Peltoniemi 2023) - is still incomplete and the scaling in the European Single Market still requires adapting to national frameworks. Appropriability conditions within the European Single Market are affected by the fact that there are both national and European layers of IPR legislation and companies must navigate this complex and uncertain environment.

To our knowledge, the emerging literature on scaling (Bohan et al. 2024, Coviello et al. 2024) does not systematically consider the role of national, European and international IPR and standardization systems and their boundaries in the scaling process. National IPR and standardization systems are institutional discontinuities in scaling processes when companies internationalize and enter new markets. Clearly, companies who are leaders in their industries and ecosystems, have strategic approach to both IPR and standardization matters, and they may "set the agenda for issues to be discussed in a standards committee" (Foss et al. 2023). Presumably, these innovation and standardization ecosystem leaders have gone through the gradual experiential learning process illustrated in Figure 3.

Figure A.2 in the online Appendix illustrates the difference in the international scaling process between born-EU firms and non-EU firms: born-EU companies must learn the functioning and evolution of national and European IPR and standards institutions and landscapes while, in contrast, non-EU firms may first scale their businesses in their domestic national markets before entering the EU Single Market and related institutional IPR and standards environment. Almost all the interviewed companies that we interviewed were originally

non-EU firms and established before Finland joined the EU in 1995. In other words, from their perspective the concept of "domestic market" can be understood in different ways: narrowly as Finland or broadly as the European Single Market.

Our study has illustrated the evolving European Single Market as a specific institutional environment which changes companies must continuously monitor (sensing), exploit (seizing) and – depending on available resources – try to impact to promote their businesses (transforming). The companies operate concurrently in multiple layered institutional environments and must play by the rules of the EU as well as the national markets in which they operate. This has important implications for studies analyzing and comparing the quality of institutions (such as appropriability conditions or regimes, e.g., patent systems) as well as their impacts on companies: analyses of national institutions and European institutions are incomplete if they do not take into account the interplay between these regional levels.

## 5.3 Managerial and policy implications

Institutional change, such as European integration, is an evolutionary process that shapes the appropriability conditions of companies (Hurmelinna-Laukkanen & Yang 2022) as well as the level of competition (Aghion et al. 2015). Companies that wish to succeed must proactively monitor and participate, if necessary, the standardization efforts. Particularly, the capability to impact the rate and direction of European integration requires investments in active participation in standardization work (Kallestrup 2017). Strategic foresight (Haarhaus & Liening 2020, Ho & Sullivan 2017) including regulatory foresight (Blind 2008)[14] and anticipation of standards seem to be key dynamic capabilities as the European integration furthers. Previously, Ho and Sullivan (2017) have emphasized the role of roadmapping standardization by governments in supporting innovation. In the context of current geopolitical environment (Zúñiga et al., 2024, Blind 2025), there is an increased risk of trade wars where IPR and standards institutions can be used as weapons of protectionism.

Regarding IPR matters, multiple interviewed companies were concerned about their FTO (cf. Guellec et al. 2012, Cappelli et al. 2023) as some competitors had received patents that – in their opinion – were non-novel and should not have been granted in the first place. As IPR landscaping software are becoming increasingly sophisticated and IPR landscaping and monitoring can be automated, companies should consider these options to support IPR strategies. Since the research interviews were conducted in 2023, the Unitary Patent system has entered into force and, as a consequence, the number of patents in force in Finland has increased heavily creating new challenges to freedom to operate analyses.

Companies were concerned about the increasing number of standards as it increases the compliance costs (e.g., need for certificates). On the other hand, this is not surprising as Bradford (2020, p. 10) notes that "corporate interests would typically advocate for a less-burdensome regulatory environment, citing costs on innovation and their international competitiveness." Still, upward harmonization leads presumably to increased compliance and adjustments costs and may therefore lead to market entry barriers (particularly for new entrants) that may help or hinder business. Recently, the European Commission has introduced several new Acts and Directives (e.g., Corporate Sustainability Reporting Directive (CSRD), Data Act, AI Act) that impact the businesses of companies operating in the European Single Market.

Cybersecurity and sustainability standards were raised by several companies as current pressing challenges. Complying with minimum cybersecurity and sustainability requirements was considered a "must have" and exceeding the minimum standards was considered to potentially promote competitiveness. Data was seen as a crucial intangible asset that enables new business opportunities. All these topics are related to concurrent European regulation and standardization developments. For instance, during the interview round, the CSRD entered into force on 5 Jan 2023.

---

[14] Blind (2008) considers standards and standardization to be elements of the regulatory framework.

Thus, policy makers could evaluate whether there is enough awareness building about these topics. In the interviews, industry associations were mentioned a couple of times as important information sources regarding standards. Already, de Vries et al. (2009) emphasized that "in improving the situation for SMEs, the role of trade associations is crucial." It seems that trade associations play an important role as knowledge brokers and supporters of companies' absorptive capacity in the context of standards and regulatory foresight. More research is needed on how trade and industry associations promote regulatory foresight and anticipation of standards and promote their member companies' capability to adapt to the evolving business environment. How large role do the trade and industry associations play in promoting regional competitiveness across countries and within the European Single Market?

Bradford (2015, p. 29) noted that "Thus, at times, the high costs of complying with the stringent EU standard lead companies to abandon the EU market altogether, instead of adjusting their global production to the EU standard." In the context of local EU companies, like our sample companies, such an option is not feasible. Local "born EU" companies must comply with the local more or less stringent standards, and this may lead to situation where they then lose bids to other companies outside the EU in cases where customers are satisfied with the lower standards. Finland has recently introduced the revised national IPR strategy and the current government (as of September 2023) is set to develop the first national standardization strategy. The former emphasizes the importance of IPR education and presumably promotion of standardization know-how and education will be an important aspect in the latter. As the interviews demonstrated, most companies had developed their current IPR and standardization approaches via learning-by-doing. Thus, more investments in IPR and standardization education would potentially promote the capability of companies to act more proactively in the evolving IPR and standardization environments. More research is needed on the role of standardization education in promoting regional competitiveness in the European Single Market and elsewhere.

## 5.4 Limitations

The article has several limitations. First, the sample is relatively small and biased towards larger and older firms which limits the generalizability of our observations. We did not provide any evidence from small and young "born EU" companies. If the evolving regulatory framework is complex, then experienced incumbent players might be better in adapting to it. On the other hand, in the presence of (suddenly) changing rules of the game, new and small more agile players might be better positioned to benefit from them if it takes longer for the larger and established firms with existing best practices to adapt their activities.

Second, there is no common definition for what is meant by "IPR strategy" or "standardization strategy". For instance, some companies focused more on standards development whereas most focused relatively more on how they implement and conform to standards (see Section 4.1). Nevertheless, the nature of the present analysis is exploratory, so despite the limited generalizability, the findings provide rich preliminary qualitative evidence.

While dynamic capabilities theory is about change over time, our semi-structured interviews offer only cross-sectional evidence. This is clearly a drawback, but it provides a big picture of the status of dynamic capabilities of regionally large companies that have years of experience. Laaksonen and Peltoniemi (2018, p. 197) note that "measurement of longitudinal data should have an important role in all types of operationalizations of dynamic capabilities". We were unable to analyze how at the time when Finland joined the EU in 1995, companies adapted to the changing environment as most of the interviewed persons did not have own experience of that era working for the company they represented. Longitudinal case studies could reveal interesting details about companies' learning patterns and sequences related to IPR and standardization strategies and dynamic capabilities development.

To conclude, more systematic identification and documentation of the sequence of learning events related to the evolving and complex IPR and standardization environment during a firm's growth, internationalization and scaling processes may provide useful lessons for new entrants. IPR and standardization institutions will remain at the core of deepening the integration in the European Single Market, and other regional innovation systems, and empirical evidence is needed to promote further evidence-based development of these systems.

## 6 Conclusions

We have shed light on the microfoundations of IPR and standardization strategies of companies operating in a peripheral and small open economy context in the European Single Market. Thus, the study contributes to the broader literatures focusing on the microfoundations of dynamic capabilities (Teece 2007, Helfat & Peteraf 2015, Felin et al. 2015) as well as experiential learning related to firm internationalization (Vahlne & Johanson 2017, Forsgren 2002, Coviello et al. 2017, Niittymies 2020).

We find that only a few of the companies have explicit IPR and standardization strategies, but almost all of them have systematic approaches to monitoring the development of standards and the IPR environments in their industries. The interviews suggest that in all cases standards play a role in defining the business environment of the companies. IPR management is to a large extent outsourced to patent attorney firms.

Most of the interviewed companies noted that their IPR and standardization strategies have evolved via "learning-by-doing" in customer projects. Since most of the interviewed companies were B2B companies, they noted that their customers and their specific requirements were important drivers of their standardization strategies. As demonstrated in this study, granular analysis of companies' experiential learning processes reveals specific learning events via which companies develop their IPR and standardization strategies and related dynamic capabilities. For policymakers, it is important to evaluate whether there is enough support for IPR and standardization education in an institutional environment where changes in IPR and standardization institutions shape appropriability conditions. How can companies shift from passive observers and reactive standards-takers towards strategic and proactive standards-developers?

Several interviewees noted that the European integration, European standards and extension of the EU have promoted their exports and international scaling of businesses. As the European integration deepens, multiple interviewed experts indicated that regulatory foresight regarding the future developments is crucial for them. On the other hand, concurrently regulatory burden and increasing compliance costs (e.g., cybersecurity and sustainability reporting requirements) were also viewed as significant challenges. Future research could study how (regulatory) foresight related to standardization and IPR institutions as a dynamic capability could promote regional competitiveness and regional innovation ecosystems.

# Appendix

**Table A.1. Interviewee characteristics**

| Higher level industry (NACE 2-digit) | Supply chain position | Turnover 2022 | Date of interview | Length of interview (min) |
|---|---|---|---|---|
| 11 Manufacture of beverages | B2B | >100M€ | 17 Feb 2023 | 34 |
| 14 Manufacture of wearing apparel | B2C | <5M€ | 21 Oct 2022 (pilot) | 52 |
| 16 Manufacture of wood and of products of wood and cork, except furniture; manufacture of articles of straw and plaiting materials | B2B | >100M€ | 19 Apr 2023 | 35 |
| 22 Manufacture of rubber and plastic products | B2C/B2B | <5M€ | 12 May 2023 | 41 |
| 22 Manufacture of rubber and plastic products | B2B | >100M€ | 29 May 2023 | 32 |
| 25 Manufacture of fabricated metal products | B2B | >100M€ | 8 Feb 2023 | 53 |
| 27 Manufacture of electrical equipment | B2B | 50-100M€ | 9 May 2023 | 37 |
| 27 Manufacture of electrical equipment | B2B | 10-50M€ | 17 May 2023 | 43 |
| 28 Manufacture of machinery and equipment | B2B | >100M€ | 10 Feb 2023 | 28 |
| 28 Manufacture of machinery and equipment | B2B | 10-50M€ | 17 Feb 2023 | 40 |
| 28 Manufacture of machinery and equipment | B2B | >100M€ | 17 Apr 2023 | 43 |
| 28 Manufacture of machinery and equipment | B2B | 5-50M€ | 3 May 2023 | 51 |
| 28 Manufacture of machinery and equipment | B2B | >100M€ | 3 May 2023 | 47 |
| 46 Wholesale trade, except of motor vehicles and motorcycles | B2C/B2B | >100M€ | 16 Jan 2023 | 39 |
| 46 Wholesale trade, except of motor vehicles and motorcycles | B2B/B2C | 10-50M€ | 30 Jan 2023 | 44 |
| 47 Retail trade, except of motor vehicles and motorcycles | B2C/B2B | >100M€ | 27 Jan 2023 | 53 |
| 62 Computer programming, consultancy and related activities | B2B | >100M€ | 28 Apr 2023 | 33 |

Notes: Source of industry is the Finnish Trade Register.

# Online Appendix / Supplemental Material

## European IPR and standards institutions

At the European level there are three European Standardization Organizations (ESOs): CEN, CENELEC and ETSI. Table A.2 illustrates the hierarchies of national, regional and international standards development organizations. Finland joined the EU in 1995, a short time after the CE-marking was introduced in 1993 (Ballor 2022, Bradford 2020). The CE ("conformité européenne") marking, means that the manufacturer or importer affirms that her goods conform to the European health, safety, and environmental protection standards. So, Finnish companies have been benefitting from the start of the CE marking. An important trend related to European integration has been the proliferation of European harmonized standards (hENs) that are developed by the aforementioned recognised European Standards Organisations following requests from the European Commission.

**Table A.2. Standardization organizations by fields and regional levels (founding years)**

| | | Field | |
|---|---|---|---|
| | **General** | **Electrotechnical** | **Telecommunication** |
| **International** | ISO (1947) | IEC (1906) | ITU (1865) |
| **European** | CEN (1961) | CENELEC (1973) | ETSI (1988) |
| **National** | SFS (1924/1947) | SESKO (1926/1965) | Traficom |

*Regional level* (vertical label on left)

**Figure A.1. Evolution of European IPR institutions and European harmonised standards**

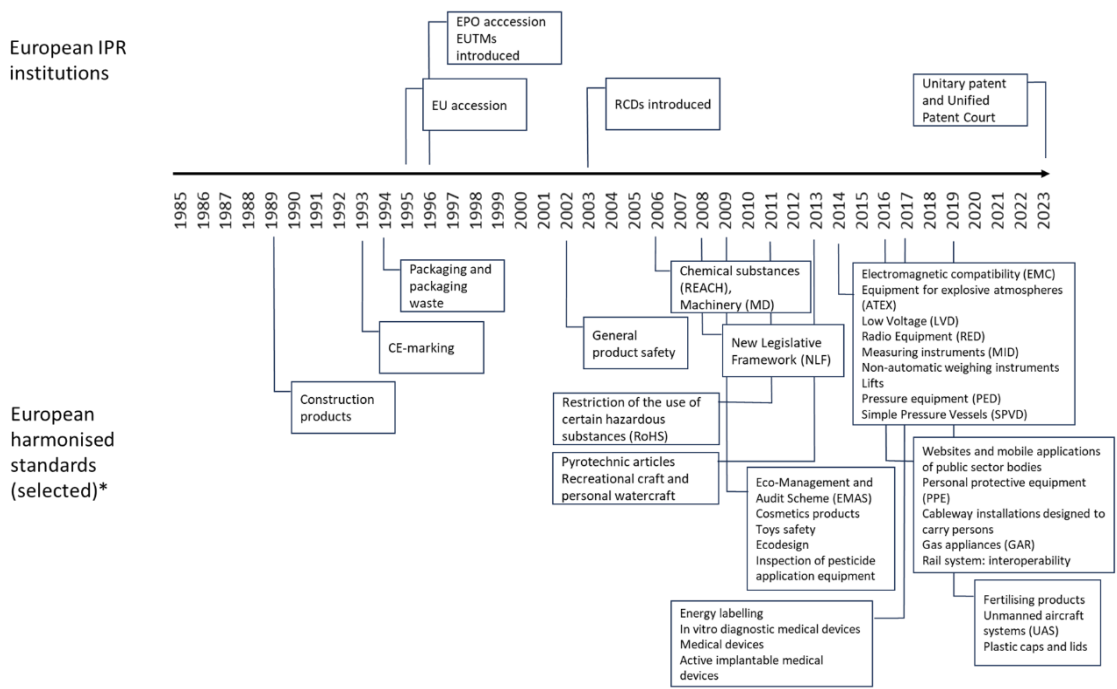

Notes: Authors' illustration. Years referring to the introduction of hENs are based on the years in which they were published in the Official Journal of the European Union. For the sake of clarity, no repealed older directives and regulations included. *The figure does not include all hENs, see full list: https://single-market-economy.ec.europa.eu/single-market/european-standards/harmonised-standards_en Last accessed 11 Oct 2023.

**Figure A.2. Complexity of institutional environments and experiential learning paths in international scaling**

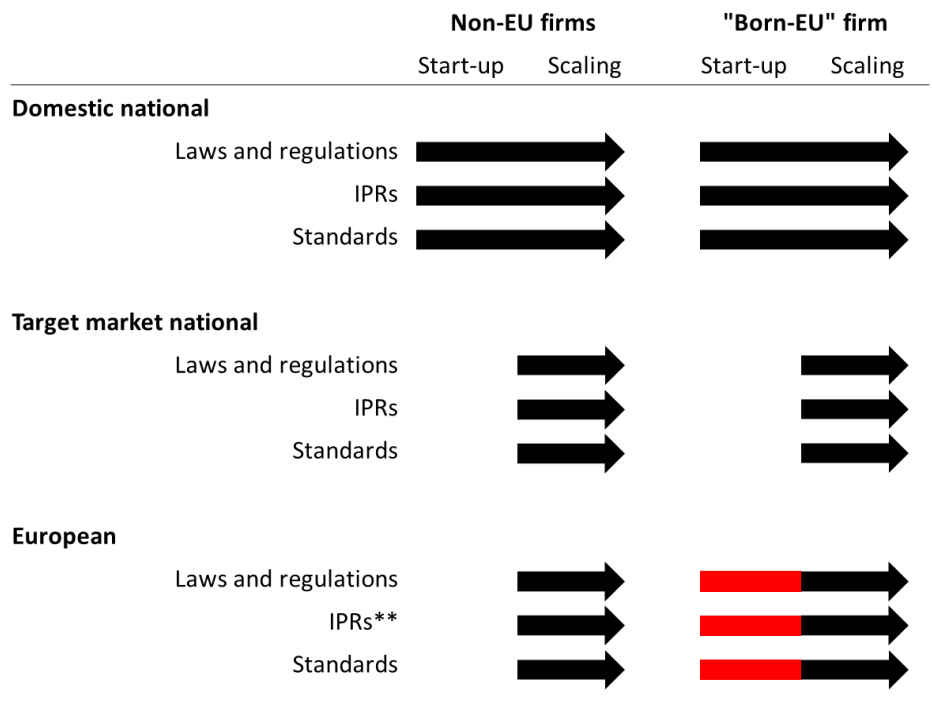

Notes: Authors' illustration. Arrows illustrate the timing of experiential learning during internationalization process of companies. **EUIPO grants European designs and trademarks and EPO, which is not an EU institution, grants European patents.

## Päijät-Häme region

Päijät-Häme (NUTS3: FI1C3) is an old industrial region (cf. Birch et al. 2010) in the southern part of Finland and city of Lahti is its capital (see e.g., Harmaakorpi & Rinkinen 2020, Aula & Harmaakorpi 2008). The distance from Brussels to Lahti is ca. 1700km, so Lahti can be considered to be a Northern peripheral area from the point of view of central Europe (cf. Spiekermann & Aalbu 2004). Historically, furniture and textile manufacturing sectors were strong pillars in the development of the region, but due to globalization and intensified international competition these industries have been hit hard during the past decades. Particularly, the great depression in early 1990s and the concurrent collapse of the Soviet Union had a long-term negative impact on local development in Lahti and Päijät-Häme region.

In developing and supporting the region's innovation ecosystem, the focus has been on practice-based innovation processes and network-facilitating innovation policy (Harmaakorpi & Rinkinen, 2020; Aula & Harmaakorpi, 2008; Pekkarinen & Harmaakorpi, 2006). The long-term strategic goal has been both building competitive resource configurations based on existing regional areas of expertise as well as economic transformation and modernization of traditional industries (Harmaakorpi, 2006; Regional Council of Päijät-Häme, 2022). The current thematic spearheads of development in the region are sport, food and beverage, and manufacturing, enhanced by sustainability as a cross-cutting theme (Regional Council of Päijät-Häme, 2022). The region has a long history in the field of environmental expertise and cleantech, and the city of Lahti was awarded as the European Green Capital 2021 by the European Commission (City of Lahti, 2021).

Konsti-Laakso et al. (2019) analyzed the IPR filing activities of local companies in Päijät-Häme region and document that local family business groups play an important role in the region as active IPR (patent, utility model, design and trademark) applicants. Hence, the Päijät-Häme region provides an interesting context for an exploratory case study of an old industrial region that is located in a peripheral location from the perspective of central Europe and the Brussels effect. For Päijät-Häme region and many alike around Europe, constructing new paths for economic development requires not only developing new technologies, but also learning about e.g. firm strategies, business models and regulatory aspects (Coenen et al. 2015). Regional efforts can have only a limited effect on new path creation unless they are integrated with industry specific supra-national institutions (such as supra-national IPR institutions).

## The structure of semi-structured interviews

The original questionnaire was in Finnish and is available from the authors upon request.

**General questions**

Does your company have

A. an IPR strategy? B. a standardization strategy?

If yes: Could you please describe them.

**Standards**

How does standards development impact your industry and business?

What are the most important standards in your field?

> A. National B. European C. International/Global

Does your company actively participate standards development? If yes: in which organizations?

> A. National B. European C. International/Global

How do you search for information about standards and monitor standards development in your field?

> A. National B. European C. International/Global

How do you develop standardization know-how in your company? (trainings, etc.)

What kind of challenges and opportunities are related to standards development in your field?

**IPR**

Which protection methods/IPRs are important in your field?

What is the importance of the following protection methods in your field?

A. Patents
B. Utility models
C. Trademarks
D. Design rights
E. Copyright
F. Something else? (e.g., data, databases)

What kind of regional and target market differences there are in the use of protection methods?

How do you search information about IPRs and monitor the development of IPR environment in your field?

To what extent do you manage IPR in-house and to what extent do you outsource? ("make or buy")

How do you develop IPR know-how in your company? (trainings, etc.)

What kind of challenges and opportunities are related to IPRs in your field?